\documentclass[conference]{IEEEtran}

\IEEEoverridecommandlockouts

\usepackage[T1]{fontenc}
\usepackage{amssymb,amsmath,epsfig}
\usepackage[latin5]{inputenc}
\usepackage{graphicx,times,latexsym}
\usepackage{psfrag,cite}
\usepackage{setspace}
\usepackage{algorithmic, algorithm}

\begin{document}

\IEEEpubid{\makebox[\columnwidth]{978-1-4799-3083-8/14/\$31.00 ©2014 IEEE\hfill }
\hspace{\columnsep}\makebox[\columnwidth]{}}

\title{A Low-Complexity Graph-Based LMMSE Receiver Designed for Colored Noise Induced by FTN-Signaling}


\author{\IEEEauthorblockN{P\i nar \c{S}en}
\IEEEauthorblockA{Electrical and Electronics Engineering\\
Middle East Technical University\\
Ankara, Turkey \\
Email: psen@metu.edu.tr}
\and
\IEEEauthorblockN{Tu\u{g}can Akta\c{s}}
\IEEEauthorblockA{Electrical and Electronics Engineering\\
Middle East Technical University\\
Ankara, Turkey \\
Email: taktas@metu.edu.tr}
\and
\IEEEauthorblockN{A. \"{O}zg\"{u}r Y\i lmaz}
\IEEEauthorblockA{Electrical and Electronics Engineering\\
Middle East Technical University\\
Ankara, Turkey \\
Email: aoyilmaz@metu.edu.tr}}

\maketitle

\begin{abstract}
We propose a low complexity graph-based linear minimum mean square error (LMMSE) equalizer which considers both the intersymbol interference (ISI) and the effect of non-white noise inherent in Faster-than-Nyquist (FTN) signaling. In order to incorporate the statistics of noise signal into the factor graph over which the LMMSE algorithm is implemented, we suggest a method that models it as an autoregressive (AR) process. Furthermore, we develop a new mechanism for exchange of information between the proposed equalizer and the channel decoder through turbo iterations. Based on these improvements, we show that the proposed low complexity receiver structure performs close to the optimal decoder operating in ISI-free ideal scenario without FTN signaling through simulations.

\end{abstract}



%
\IEEEpeerreviewmaketitle

\IEEEpubidadjcol

\section{INTRODUCTION}
\label{sec:intro}
There are some proposed techniques to increase the spectral efficiency in the literature, one of which is the Faster-than-Nyquist (FTN) signaling. The first studies on FTN signaling concept date back to 1970s~\cite{Mazo1975}. But it has received much more attraction recently as a means of providing higher transmission rate beyond the Nyquist criterion in the same spectral shape consuming the same energy per bit~\cite{Rusek2013}. By contrast to the classical scenario using $T$-orthogonal pulse shape, in FTN signaling, the pulses can be packed by violating the Nyquist rate without decreasing the minimum Euclidean distance ($d_{min}^2$) in the signaling space~\cite{Rusek2006}. The minimum symbol time until which $d_{min}^2$ is not below the value of the case with orthogonal pulse shape is called the Mazo limit~\cite{Rusek2007}. 

Since FTN signaling has more symbols to be packed in the time interval $T$ than the conventional orthogonal signaling, there exists intentional intersymbol interference (ISI) which causes an increase in the receiver complexity. However, thanks to the recent studies on practical receivers, it is still possible to achieve the same error rate performance as the conventional way. Among the latest ones, a reduced trellis based algorithm ($M$-BCJR) having a linearly increasing complexity with block length is proposed in~\cite{Anderson2012}. But its complexity increases exponentially with constellation size and the number of ISI taps due to the necessity of higher $M$ value. Moreover, $M$-BCJR needs an optimized whitening filter for each channel realization at the receiver side so as to give a better performance as mentioned in~\cite{Anderson2012} which makes this method hard to implement in real time for fading environments. In another work~\cite{Sugiura2013}, frequency domain equalization with an additional complexity of fast fourier transform (FFT) and inverse-FFT operations is analysed for uncoded FTN schemes. It brings on a performance loss because of the lack of coding and turbo operation. However, what we propose in this paper is a low complexity (linearly increasing with block length) and practical reduced LMMSE equalization method to remedy the ISI effect due to FTN signaling. Furthermore, our receiver structure is perfectly suited for high constellation sizes, since the number of the elements in the alphabet is irrelevant to our equalization process. In a more detailed way, we develop a factor graph-based LMMSE algorithm in which the non-white noise inherent in FTN is taken into consideration. The proposed LMMSE equalizer operates on the input signal with the assumption that it is coming from a Gaussian alphabet~\cite{Loeliger2006}. In addition, based on this Gaussian approximation, a new calculation technique in extrinsic information exchange mechanism is suggested for turbo operations. The receiver structure that we propose is shown to perform very close to the optimal decoder operating under non-ISI, i.e., additive white Gaussian noise (AWGN) channel through extensive simulations.

The notations used in the paper are organized as follows. Lower case letters (e.g. $x$) denote scalars, lower case bold letters (e.g. $\mathbf{x}$) denote vectors, upper case bold letters (e.g. $\mathbf{X}$) denote matrices. For a given random variable $x$; $m_x$, $v_x$ and $w_x$ denote its mean, variance and weight respectively where $w_x=v_x^{-1}$. For a given vector random variable $\mathbf{x}$; $\mathbf{R_x},\mathbf{m_x},\mathbf{V_x},\mathbf{W_x}$ denote 
its autocorrelation matrix, mean vector, covariance matrix and weight matrix respectively where $\mathbf{W_x}=\mathbf{V_x}^{-1}$. The indicators $()^*$, $()^H$, and $E\{\}$ denote conjugate, hermitian transpose and expectation operations respectively.

The paper is organized as follows. Section~\ref{sec:system_model} presents the system model and autoregressive (AR) process approximation for the noise signal. Section~\ref{sec:lmmse_graphs} presents factor graph-based LMMSE receivers including our reduced complexity receiver design. In secton~\ref{sec:Simulation}, bit error rate (BER) performance results of our developed receiver structure are given. Lastly, Section~\ref{sec:conclsn} concludes the paper.
\section{SYSTEM MODEL}
\label{sec:system_model}
\subsection{Signal Model}
\label{sec:signal_model}
We consider a baseband communication system using $\mathcal{M}$-point complex valued phase shift keying ($\mathcal{M}$-PSK) or quadrature amplitude ($\mathcal{M}$-QAM) modulation from an alphabet $S$. Fig.~\ref{fig:system_model_ftn} presents the whole block diagram of the transmitter and the receiver structures. As given in Fig.~\ref{fig:system_model_ftn}, at the transmitter side, an information symbol $x_m$ which is modulated from $l$ coded information bits is passed through a $T$-orthogonal root-raised-cosine (rRC) pulse shaping filter $p(t)$, where $m$ is the symbol index. The average symbol energy is $E_s$, i.e., $E_s=E\lbrace{\vert{x_m}\vert}^2\rbrace$. The signaling rate is $1/\tau T$; in other words, each symbol is transmitted in a time duration of $\tau T$. Here, $\tau$ is the packing ratio of symbols which ranges between $0$ and $1$. If $\tau=1$, system does not suffer from ISI since the pulse is T-orthogonal. ISI pattern arising from FTN signaling occurs when $\tau<1$ and as $\tau$ decreases, transmission rate increases.  $\alpha$ is the excess bandwidth of rRC filter and we assume $p(t)$ is of unit energy. The received  signal under the AWGN channel is written as follows~\cite{Rusek2006}:
\begin{figure}[htbp]
   \centering
   \includegraphics[width=0.5\textwidth]{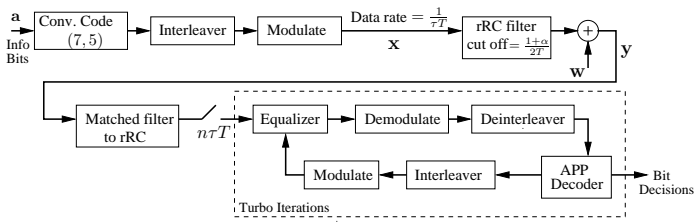}
   \caption{System Model}
   \label{fig:system_model_ftn}
\end{figure}
\begin{align}
\label{eqn:observation}
y(t) = \sum_{m=-\infty}^\infty x_m p(t-m\tau T) + w(t),   \tau <1
\end{align}
where $w(t)$ represents aditive white zero mean circularly symmetric complex Gaussian noise with total variance $N_0$, i.e., $w(t) \sim CN(0,N_0)$. The signal at the $k$'th discrete time instance at the output of the matched filter after sampling with $\tau T$ will become
\begin{align}
\label{eqn:after_sampling}
r_k = \int_{-\infty}^{\infty} y(t)p^*(t-k\tau T)dt.
\end{align}
Inserting the expression for $y(t)$ in (\ref{eqn:observation}) gives
\begin{align}
\label{eqn:receive_discrete}
r_k = \sum_{m=-\infty}^\infty x_m h[m-k] + n_k,
\end{align}
where
\begin{align}
\label{eqn:chan_coeff}
h[m-k] = \int_{-\infty}^{\infty} p(t-m\tau T)p^*(t-k\tau T)dt
\end{align}
and
\begin{align}
\label{eqn:noise}
n_k = \int_{-\infty}^{\infty} w(t)p^*(t-k\tau T)dt.
\end{align}
The autocorrelation of the noise sequence $\mathbf{n}$ is then
\begin{align}
\label{eqn:noise_autocorr}
E\left\lbrace n[m]n^*[k]\right\rbrace = N_0 h[m-k].
\end{align}
As mentioned above, FTN results in unavoidable ISI and non-white noise samples which in turn worsen the error performance and increase the receiver complexity. What we propose to overcome these issues is a factor graph based LMMSE equalization method in which the non-white noise is taken into consideration while obtaining higher transmission rate with higher performance. Our proposed equalizer is thought to be used in a turbo operation together with an \textit{a posteriori} probability (APP) decoder as given in Fig.~\ref{fig:system_model_ftn}.

Before getting into details of the equalizer, a short introduction to the AR process  which is used for modelling the non-white noise is given in the proceeding section.
\subsection{AR Process Model for Colored Noise}
\label{sec:ar_process}
An AR process of order $p$ is a stochastic process which can be described by the weighted sum of its $p$ previous values and a white innovation term as defined in \cite{Bhat1972}. The value of $p$-order AR process at time $k$ is given in (\ref{eqn:ar_process}) where $\tilde{n}_k$ is the white innovation term at time $k$ and $\psi_i$'s are the AR parameters for $i=0,1,\ldots,p$.
\begin{align}
\label{eqn:ar_process}
n_k = \sum_{i=1}^{p}\psi_i n_{k-i} + \tilde{n}_k.
\end{align}
The autocorrelation function of an AR sequence, $\gamma(j)$, with known parameters ($\psi_i$'s) and the covariance of innovation term ($\sigma_{\tilde{n}}^2$) can be found by the Yule-Walker equations as given in \cite{Bhat1972}:
\begin{equation}
\label{eqn:ar_corr}
\gamma(j) = 
\begin{cases}
\sum_{i=1}^{p}\psi_i \gamma(-i) + \sigma_{\tilde{n}}^2 & \text{for } j=0  \\
\sum_{i=1}^{p}\psi_i \gamma(j-i) & \text{for }j>0.
\end{cases}
\end{equation}
Conversely, if the first $p+1$ values of autocorrelation function are known, one can reach $\psi_i$ for $i=0,1,\ldots,p$ and $\sigma_{\tilde{n}}^2$ through the Yule-Walker equations. In our problem, non-white Gaussian noise is a moving average process. However, we can still approximate it with a $p$-order AR process by choosing a proper value for $p$. We propose such an approximation method for constructing the LMMSE equalizer structure which is detailed in Section~\ref{sec:red_i_lmmse}.
\section{GRAPH BASED LMMSE EQUALIZER}
\label{sec:lmmse_graphs}
\subsection{Iterative LMMSE Equalizer (I-LMMSE)}
\label{sec:i_lmmse}
\begin{figure}[htbp]
   \centering
   \includegraphics[width=0.47\textwidth]{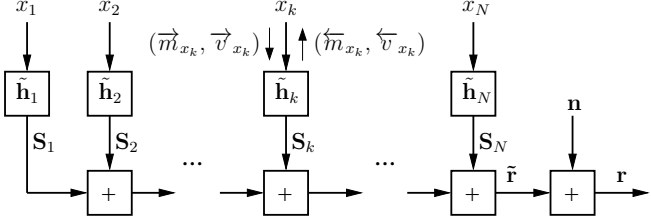}
   \caption{Factor Graph of I-LMMSE}
   \label{fig:i_lmmse_graph}
\end{figure}

The system described by (\ref{eqn:observation}) can be rewritten as in (\ref{eqn:observation_matrix}) where $\mathbf{H}$ is the convolution matrix of size $N \times N$ originating from the ISI pattern and $N$ is the block length.
\begin{align}
\label{eqn:observation_matrix}
\mathbf{r} =& \mathbf{H} \mathbf{x} + \mathbf{n} ,
\end{align}
\begin{align}
\label{eqn:conv_matrix}
\mathbf{H} = \left[ \begin{array}{cccccc}
h_0 &\,h_{-1} & \ldots & h_{-L} & & \\
\vdots &\,h_0 & & \quad\quad\ddots  & & \\
h_L &\,\ldots & h_0 & \ldots & h_{-L} &  \\
&\, & & & & \\
&\, h_L & \ldots & h_0 &  & h_{-L} \\
&\, & \ddots &  &  & \vdots \\
&\, &  & h_L & \ldots & h_0 \\
 \end{array} \right].
\end{align}   
Previously proposed factor graph of the I-LMMSE equalizer for a system given in equation (\ref{eqn:observation_matrix}) is shown in Fig.~\ref{fig:i_lmmse_graph} \cite{Loeliger2007}. In this generic graph, $\tilde{\mathbf{h}}_i$'s ($i=1,2,\ldots ,N$) are the columns of $\mathbf{H}$ and $\mathbf{n}$ is the noise vector of length $N$. Symbols are represented by mean and variance values in this graph which implements the block MMSE filtering operation through Gaussian message passing rules in \cite{Loeliger2007}. Assuming channel coded operation, I-LMMSE equalizer carries out its calculations in an iterative way as shown in Fig.~\ref{fig:system_model_ftn} by means of the \textit{a priori} information coming from the APP decoder. 
The aim is to compute the \textit{a posteriori} mean and variance values of the symbols (${m}_{x_i}^{post}$,\:${v}_{x_i}^{post}$ for $i=1,2,\ldots,N$) by combining the incoming messages ($\overrightarrow{m}_{x_i}$,$\overrightarrow{v}_{x_i}$) obtained from the output of the APP decoder and the outgoing messages ($\overleftarrow{m}_{x_i}$,$\overleftarrow{v}_{x_i}$) as follows~\cite{Loeliger2006}:
\begin{align}
\label{eqn:message_ilmmse_var_apos}
{v}_{x_i}^{post}=&(\overrightarrow{v}_{x_i}^{-1}+\overleftarrow{v}_{x_i}^{-1})^{-1}, 
\end{align}
\begin{align}
\label{eqn:message_ilmmse_mean_apos}
{m}_{x_i}^{post}=&{v}_{x_i}^{post}(\overrightarrow{v}_{x_i}^{-1} \overrightarrow{m}_{x_i}+\overleftarrow{v}_{x_i}^{-1} \overleftarrow{m}_{x_i}).
\end{align}
One way to reach $\overleftarrow{m}_{x_i}$ and $\overleftarrow{v}_{x_i}$ by using the observation vector $\mathbf{r}$, the incoming messages and the statistical information of noise signal through the message passing rules~\cite{Loeliger2007} is below where $\tilde{\mathbf{W}}_\mathbf{x}$ represents the auxiliary quantity defined in~\cite{Loeliger2007}:
\begin{align}
\label{eqn:message_ilmmse_var}
\tilde{\mathbf{W}}_{\mathbf{\tilde{r}}} \triangleq  & \left(\overrightarrow{\mathbf{V}}_{\mathbf{\tilde{r}}}+\overleftarrow{\mathbf{V}}_{\mathbf{\tilde{r}}}\right)^{-1}     \\
\tilde{\mathbf{W}}_{\mathbf{\tilde{r}}} =  & \tilde{\mathbf{W}}_{\mathbf{S}_i} \text{   for } i=1,2,\ldots ,N \\
\overleftarrow{v}_{x_i} = & {\tilde{w}_{{x}_i}}^{-1} - \overrightarrow{v}_{x_i} \\
=  & \left({\tilde{\mathbf{h}}_i^H \tilde{\mathbf{W}}_{\mathbf{\tilde{r}}}} \tilde{\mathbf{h}}_i \right)^{-1}-\overrightarrow{v}_{x_i}
\end{align} 
\begin{align}
\label{eqn:message_ilmmse_mean}
\overleftarrow{m}_{x_i} = & {\tilde{w}_{{x}_i}}^{-1} \tilde{\mathbf{h}}_i^H \tilde{\mathbf{W}}_{\mathbf{\tilde{r}}} \overleftarrow{\mathbf{m}}_{\mathbf{S}_i}\\
=  & \left({\tilde{\mathbf{h}}_i^H \tilde{\mathbf{W}}_{\mathbf{\tilde{r}}}} \tilde{\mathbf{h}}_i \right)^{-1} \tilde{\mathbf{h}}_i^H \tilde{\mathbf{W}}_{\mathbf{\tilde{r}}} \overleftarrow{\mathbf{m}}_{\mathbf{S}_i}
\end{align} 
where,
\begin{align}
\label{eqn:message_ilmmse_mean_2}
\overleftarrow{\mathbf{m}}_{\mathbf{S}_i} = & \overleftarrow{\mathbf{m}}_{\tilde{\mathbf{r}}} - \sum_{l=1}^{N} {\tilde{\mathbf{h}}_l \overrightarrow{m}_{x_l}} + \tilde{\mathbf{h}}_i \overrightarrow{m}_{x_i}
\end{align} 
We have zero mean noise vector with autocorrelation $\mathbf{R_n}$ which results in $\overleftarrow{\mathbf{V}}_{\mathbf{\tilde{r}}} = \mathbf{R_n}$ and $\overleftarrow{\mathbf{m}}_{\mathbf{\tilde{r}}} = \mathbf{r}$ in equations (\ref{eqn:message_ilmmse_var})-(\ref{eqn:message_ilmmse_mean_2}). Since I-LMMSE processes vectors and matrices of size $N$, this approach has a computational complexity of order $O(N^2)$. On the other hand, its complexity is not affected by the constellation size $\mathcal{M}$. 
\subsection{Reduced Iterative LMMSE Equalizer (RI-LMMSE)} 
\label{sec:red_i_lmmse}
It is shown that the factor graph in Fig. \ref{fig:i_lmmse_graph} can be separated element-wise under white Gaussian noise signal operation which results in a linearly increasing complexity with $N$ in \cite{Guo2008}. However, no work on factor graphs in which the non-white noise is taken into consideration  exists in the literature within the knowledge of the authors. On the other hand, the Kalman filtering operation is discussed under AR-$p$ process modelled (non-white) Gaussian noise in \cite{Gibson1991} from a signal processing point of view with no emphasis on equalization. The main idea of \cite{Gibson1991} is to concatenate the set of state variables and the set of noise variables of length $p$. Hence, we propose a new method based on \cite{Gibson1991} and \cite{Guo2008} so as to implement the LMMSE filtering operation on a factor graph which has a linearly increasing complexity with $N$. In our factor graph representation, it is assumed that the system model is described by the equation (\ref{eqn:observation_matrix}) and the noise term results from a $p$-order AR process. Let $\mathbf{h}=[h_{L}\;h_{L-1}\;\ldots\;h_0\;\ldots\;h_{-L}]$ denote the ISI pattern coefficient vector and ${\Psi}=[\psi_p\;\psi_{p-1}\;\ldots\;\psi_1]$ denote the AR process parameter vector. Then the $k^{\textit{th}}$ element of the observation vector $\mathbf{r}$ can be rewritten as
\begin{align}
\label{eqn:observation_ar_noise}
r_k = \mathbf{h} \; [x_{k-L}\;x_{k-L+1}\;\ldots\;x_k\;\ldots\;x_{k+L}]^T + n_k,
\end{align} 
where
\begin{align}
\label{eqn:ar_noise}
n_k = {\Psi} \;[n_{k-p}\;n_{k-p+1}\;\ldots\;n_{k-1}]^T + \tilde{n}_k.
\end{align} 
Inserting the expression for $n_k$ in (\ref{eqn:ar_noise}) yields
\begin{align}
\label{eqn:observation_FG}
r_k = \mathbf{\overline{h}}\; \mathbf{\overline{x}}_k^T + \tilde{n}_k,
\end{align}
where $\mathbf{\overline{h}}=[\mathbf{h} \quad \Psi]$ and
\begin{align}
\label{eqn:FG_matrix_defn}
\mathbf{\overline{x}}_k = [x_{k-L}\;\ldots\;x_k\;\ldots\;x_{k+L}\;n_{k-p}\;\ldots\;n_{k-1}]^T.
\end{align} 
\begin{figure*}[t]
   \centering
   \includegraphics[width=0.7\textwidth]{figures/FTN_LMMSE}
   \caption{Factor Graph of RI-LMMSE}
   \label{fig:red_i_lmmse_graph}
\end{figure*} 
In our factor graph representation, $\mathbf{\overline{x}}_k$ denotes the joint state variables which needs to be updated within each building block. So, we define
\begin{align}
\label{eqn:FG_matrix_defn_G}
\mathbf{G} =& \left[ \begin{array}{cccc}
\mathbf{0}_{2L \times 1} & \mathbf{I}_{2L}  & \mathbf{0}_{2L \times 1} & \mathbb{O}_{2L \times (p-1)} \\
0 & \mathbf{0}_{1 \times 2L} & 0 & \mathbf{0}_{1 \times (p-1)} \\
\mathbf{0}_{(p-1) \times 1} & \mathbb{O}_{(p-1) \times 2L} & \mathbf{0}_{(p-1) \times 1} & \mathbf{I}_{p-1} \\
0 & \mathbf{0}_{1 \times 2L} & & \hskip -5em [\;\text{----------}\;\Psi\;\text{----------}\;]\\
 \end{array} \right], \\
\mathbf{F} =& \left[ \begin{array}{cc}
\mathbf{0}_{2L \times 1} & \mathbf{0}_{2L \times 1} \\
1 & 0 \\
\mathbf{0}_{(p-1) \times 1} & \mathbf{0}_{(p-1) \times 1}  \\
0 & 1 \\
 \end{array} \right]
\end{align} 
where $\mathbf{I}_J$ denotes the identity matrix of size $J \times J$, $\mathbf{0}_J$ denotes the all zero vector of length $J$ and $\mathbb{O}$ denotes the all zero matrix of the specified size. The update of the state variables occurs through the use of $\mathbf{F}$ and $\mathbf{G}$ as follows:
\begin{align}
\label{eqn:FG_state_defn_x_k1}
\mathbf{\overline{x}}_{k+1} =& \mathbf{F}\;[x_{k+L+1} \quad \tilde{n}_{k}]^T + \mathbf{y}_{k+1}, 
\end{align}
where
\begin{align}
\label{eqn:FG_state_defn_y}
\mathbf{y}_{k+1}=&\mathbf{G}\; \mathbf{\overline{x}}_{k}.
\end{align}
The operations carried out in (\ref{eqn:observation_ar_noise})-(\ref{eqn:FG_state_defn_y}) can be seen on the factor graph given in Fig.~\ref{fig:red_i_lmmse_graph}. The logic behind the function of this tree like structure is similar to that in sectin~\ref{sec:i_lmmse} such that the symbols are represented by their mean and variance values on the graph. The aim is to find the \textit{a posteriori} mean and variance values of the state variables ($\mathbf{V}_{\overline{\mathbf{x}}_k}^{post},\mathbf{m}_{\overline{\mathbf{x}}_k}^{post}$) through the process of forward and backward recursions which give the incoming and outgoing messages respectively as follows~\cite{Guo2008,Loeliger2006}:
\begin{align}
\label{eqn:message_redilmmse_var_apos}
\mathbf{V}_{\overline{\mathbf{x}}_k}^{post}=&(\overrightarrow{\mathbf{V}}_{\overline{\mathbf{x}}_k}^{-1}+\overleftarrow{\mathbf{W}}_{\overline{\mathbf{x}}_k})^{-1}, 
\end{align}
\begin{align}
\label{eqn:message_redilmmse_mean_apos}
\mathbf{m}_{\overline{\mathbf{x}}_k}^{post}=&\mathbf{V}_{\overline{\mathbf{x}}_k}^{post}(\overrightarrow{\mathbf{V}}_{\overline{\mathbf{x}}_k}^{-1} \overrightarrow{\mathbf{m}}_{\overline{\mathbf{x}}_k} +\overleftarrow{\mathbf{W}}_{\overline{\mathbf{x}}_k} \overleftarrow{\mathbf{m}}_{\overline{\mathbf{x}}_k}).
\end{align}
Corresponding operations of the blocks included in the forward and backward recursions on the factor graph in Fig.~\ref{fig:red_i_lmmse_graph} are developed in~\cite{Loeliger2007,Loeliger2006,Guo2008}. We use these operations to reach the information on the state variables from which the data symbol related part is extracted (noise part is stripped out) after performing the calculations given in (\ref{eqn:message_redilmmse_var_apos})-(\ref{eqn:message_redilmmse_mean_apos}). The \textit{a posteriori} mean and variance values of importance are indeed included in the upper part of size ($2L+1$) of the joint state variables. It can be noted that the \textit{a priori} mean and variance value for the $(k+L+1)^{\textit{th}}$ input symbol ($m_{x_{k+L+1}}^{prior}$, $v_{x_{k+L+1}}^{prior}$) get involved in the graph being concatenated with the mean and variance value of the white Gaussian noise $\tilde{n}_{k}$ as in (\ref{eqn:FG_state_defn_x_k1}). Since $x_{k+L+1}$ and $\tilde{n}_{k}$ are independent random variables, the total \textit{a priori} information given to the $k^{\textit{th}}$ building block of the graph ($\mathbf{m}^{prior}_k$,$\mathbf{v}^{prior}_k$) can be written as:
\begin{align}
\label{eqn:red_lmmse_aprior}
\mathbf{m}^{prior}_k =& \left[ \begin{array}{c}
m_{x_{k+L+1}}^{prior}\\
0
 \end{array} \right], \\
\mathbf{v}^{prior}_k =& \left[ \begin{array}{cc}
v_{x_{k+L+1}}^{prior} & 0\\
0 & \sigma^2_{\tilde{n}}
 \end{array} \right]. 
\end{align}
Next we move on to the exchange of these generated values with the ones from APP decoder.
\subsection{Information Exchange Between APP Decoder and LMMSE Equalizer}
\label{sec:llr_comm}
\begin{figure}[htbp]
   \centering
   \includegraphics[width=0.45\textwidth]{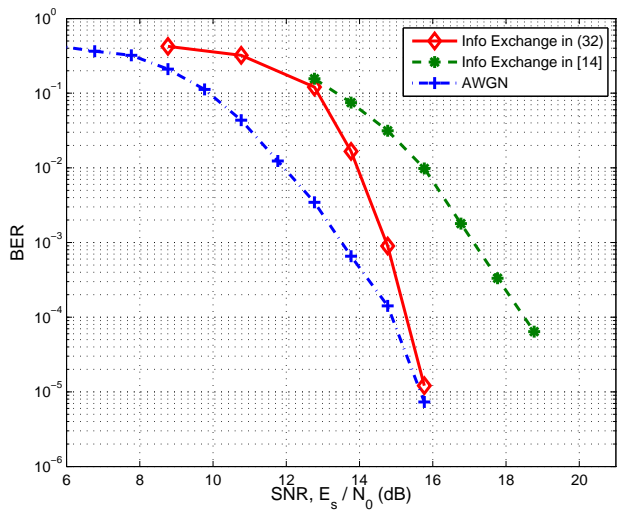}
   \caption{Performance comparison of extrinsic LLR value calculation methods}
   \label{fig:ext_llr_comp}
\end{figure}

\textit{A priori} information of the symbols ($m_{x_k}^{prior}$,$v_{x_k}^{prior}$) given to the LMMSE equalizer can be obtained by applying the Gaussian approximation to the log likelihood ratio (LLR) of the bits coming from the APP decoder. After the process of LMMSE equalization, the obtained extrinsic LLR values are passed to the APP decoder for the next iteration. An earlier method for calculation of the extrinsic LLR values utilizing the Gaussian approximation to the LMMSE filter output is proposed in~\cite{Wangpoor1999} and later used in~\cite{Tuchler2002}. However, this approach is implemented on the vectors of size $N$ which results in a complexity $O(N^2)$. In order to decrease this complexity, an equivalent method with simplified form is given in~\cite{Guo2008} for only BPSK signaling. On the other hand, for $\mathcal{M}$-QAM signaling, there is no such a simplified work for extrinsic LLR computation in the literature except~\cite{Guo2011}, which proposes the  direct subtraction of the \textit{a priori} LLR values coming from APP decoder in equation (16) in~\cite{Guo2011}. However, it causes non-ignorable performance loss since LMMSE operates under Gaussian input signal approximation. Hence, we propose a new procedure for the calculation of extrinsic LLR values, which is particularly of importance for the scenario utilizing higher constellation size. The main idea of our method is to subtract the recalculated \textit{a priori} LLR value which is generated from the Gaussian distribution with given \textit{a priori} mean and variance ($m_{x_k}^{prior}$,$v_{x_k}^{prior}$). If the modulated symbol $x_k$ is represented by the bits of $[c_{k,1}\;c_{k,2} \ldots c_{k,l}]$, then the output extrinsic LLR values ($L^{ext}$) are obtained as follows:
\begin{align}
\label{eqn:lmmse_out_extrinsic_llr}
L^{ext}(c_{k,q}) = L^{int}(c_{k,q}) - L^{prior}(c_{k,q})
\end{align}
where $L^{int}$ denotes the intrinsic LLR and $L^{prior}$ denotes the recalculated \textit{a priori} LLR values. The intrinsic LLR values in (\ref{eqn:lmmse_out_extrinsic_llr}) is obtained as follows:
\begin{align}
\label{eqn:lmmse_out_intrinsic_llr}
L^{int}(c_{k,q}) = \ln\left(\frac{\sum_{s_i\in S_{q,0}} P(x_k=s_i|\mathbf{r})}{\sum_{s_i\in S_{q,1}} P(x_k=s_i|\mathbf{r})}\right),\; q=0,1,\ldots ,l
\end{align}
where $s_i$ denotes the $i^{\textit{th}}$ element of the symbol alphabet $S$, $S_{q,0}$ and $S_{q,1}$ denote the subsets of $S$ with sequences whose $q^{\textit{th}}$ bit is $0$ and $1$ respectively. The probability of $P(x_k=s_i|\mathbf{r})$ in (\ref{eqn:lmmse_out_intrinsic_llr}) is calculated based on the conditional probability density function (pdf) of the $k^{\textit{th}}$ input symbol obtained from (\ref{eqn:message_ilmmse_var_apos})\&(\ref{eqn:message_ilmmse_mean_apos}) for I-LMMSE or (\ref{eqn:message_redilmmse_var_apos})\&(\ref{eqn:message_redilmmse_mean_apos}) for RI-LMMSE as:
\begin{align}
\label{eqn:lmmse_out_gauss}
p_{x_k}(a|\mathbf{r}) = \frac{1}{\pi v^{post}_{x_k}} e^{-\frac{|a-m^{post}_{x_k}|^2}{v^{post}_{x_k}}}.
\end{align}
The \textit{a priori} LLR values in (\ref{eqn:lmmse_out_extrinsic_llr}) is obtained by a process to (\ref{eqn:lmmse_out_intrinsic_llr}) as follows:
\begin{align}
\label{eqn:lmmse_out_apriori_llr}
L^{prior}(c_{k,q}) = \ln\left(\frac{\sum_{s_i\in S_{q,0}} P(x_k=s_i)}{\sum_{s_i\in S_{q,1}} P(x_k=s_i)}\right).
\end{align}
In (\ref{eqn:lmmse_out_apriori_llr}), the probability of $P(x_k=s_i)$ is calculated based on the fact that $x_k \sim CN(m_{x_k}^{prior},v_{x_k}^{prior})$ under Gaussian approximation. It is observed by trial and error that multiplying the LLR values exchanged between the decoder and the equalizer by different scaling factors greatly affects the BER performance. A very similar method is also applied in \cite{Colavolpe2009} for turbo equalization and decoding structures. The performance comparison between the extrinsic LLR calculation method we propose and the one in \cite{Guo2011} is shown in Fig.~\ref{fig:ext_llr_comp}. For both method, LMMSE equalizer in \cite{Guo2008} is implemented for a $6$-tap static ISI pattern of $[0.408\;0\;0\;0\;0.816\;0.408]$ which is taken from~\cite{Colavolpe2005} under white Gaussian noise with $64$-QAM signaling. The data length is set to 1800 uncoded bits. The convolutional code with code rate $1/2$ and generator polynomial $[133\;171]$ is used. The number of turbo iterations is $5$. Our method is operated with the scaling factor of $0.5$ which multiplies the extrinsic LLR values at the output of the equalizer. It is seen that at $10^{-4}$ BER level, there is more than $3$ dB gain of our method with respect to the one in~\cite{Guo2011}. Another important point to mention is that the method in~\cite{Guo2011} leads to no improvement in performance as the number of turbo iterations increases. Besides, no enhancement is observed in its performance with the help of a scaling factor through simulations.

\subsection{Demodulating Complexity Comparison}
\label{sec:complexity}
In FTN signaling, trellis based algorithms such as the VA and BCJR are frequently used in the literature due to the existence of ISI. However, the trellis structure becomes extremely large when there is a high delay spread of ISI with small $\tau$ values and/or whenever a high-order constellation is utilized. There are recently proposed reduced complexity algorithms in~\cite{Sugiura2013} and~\cite{Anderson2012}. In~\cite{Anderson2012}, $M$-BCJR algorithm based on searching only a subset of whole trellis is analysed. Its approximate complexity is $O(MN)$ with the additional complexity due to the use of a whitening filter, where $M$ is the number of trellis states visited at each symbol. Hence, the required $M$ value for sensible BER performance increases exponentially with constellation size $\mathcal{M}$ and number of ISI taps. Moreover, there needs to be an optimization of the whitening filter for the $M$-BCJR algorithm as mentioned in~\cite{Anderson2012} which may lead to a problem for fading environments with this challenging optimization necessity. In~\cite{Sugiura2013}, a frequency domain MMSE equalization method with complexity of $O\left(N\log(N)\right)$ is proposed for only uncoded FTN systems without any adaptation to coding schemes. Because of the absence of a coding scheme and a turbo iterative structure using soft information of bits, it is observed that there is 3-5 dB performance loss in BER performance even above the Mazo limit (higher $\tau$ than $\tau^*$). Furthermore, one should remember that I-LMMSE given in Section~\ref{sec:i_lmmse} may not be the first choice either due to its complexity of $O(N^2)$. However, I-LMMSE can still serve as a means to put a lower bound on our proposed RI-LMMSE since it uses the noise statistics directly without any approximations. On the other hand, by using a similar way to~\cite{Guo2008}, the complexity of RI-LMMSE in section~\ref{sec:red_i_lmmse} is $O(N\mathcal{L}^2)$ where $\mathcal{L}$ is the sum of the number of ISI taps and the parameter number ($p$) used in the approximation of the noise process to an AR process. This is because RI-LMMSE operates over the matrices of size $\mathcal{L} \times \mathcal{L}$ for each building block. It should be noted that $\mathcal{L}$ does not change with increasing constellation size and scales up only linearly with enlarging ISI pattern. Moreover, RI-LMMSE can easily be adapted to the fading environments by modifying the channel taps in the building blocks.

\section{SIMULATION RESULTS}
\label{sec:Simulation}
\begin{figure}[htbp]
   \centering
   \includegraphics[width=0.45\textwidth]{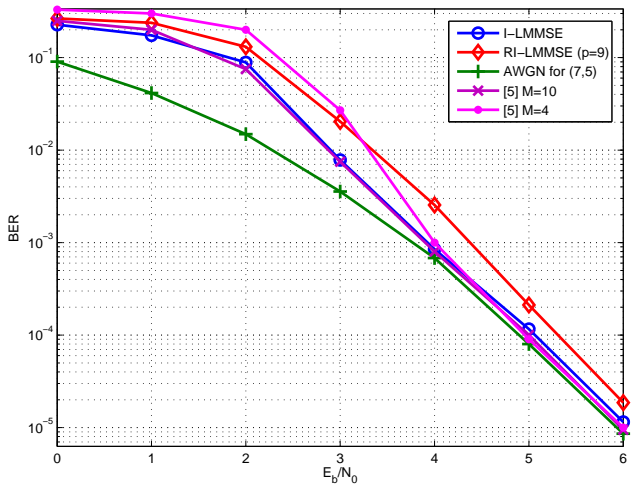}
   \caption{Performance of I-LMMSE and RI-LMMSE for FTN signaling: BPSK}
   \label{fig:FTN_bpsk}
\end{figure}

In this section, we present our performance results of FTN signaling for BPSK and $16$-QAM modulation schemes. We assume that a rRC filter with $\alpha =0.3$ and time delay of $8T$ is implemented. The convolutional code with code rate $1/2$ and generator polynomial ($7,5$) is used. The Mazo limit under that case is $1/3$~\cite{Rusek2007}. Uncoded $3000$ bits are transmitted over AWGN channel for RI-LMMSE receiver and $750$ bits for I-LMMSE receiver due to its high complexity. Simulation results for BPSK signaling under $\tau=0.5$ case are shown in Fig.~\ref{fig:FTN_bpsk} where maximum $15$ of the ISI taps take part in the RI-LMMSE equalization procedure. AR process approximation of the noise is performed according to (\ref{eqn:noise_autocorr}) with $p =9$ which is chosen by trial and error. A more detailed information regarding the selection of AR process parameters and its effects on the performance results is left as a future work. The scaling factors of ($0.5,0.5$) are chosen to multiply the LLR values of the equalizer and decoder output respectively for RI-LMMSE receiver while ($1,0.5$) are chosen for those of I-LMMSE receiver. The number of turbo iterations is $15$. It is seen from Fig.~\ref{fig:FTN_bpsk} that RI-LMMSE receiver structure we propose performs very close to no ISI case with a loss of only less than $0.3$ dB at BER of $10^{-4}$. Also, the performance of the $M$-BCJR algorithm in~\cite{Anderson2012} for the same scenario under BPSK signaling is shown in Fig.~\ref{fig:FTN_bpsk} for $M=4$ and $M=10$ values. Its performance is also close to the no ISI case.

\begin{figure}[htbp]
   \centering
   \includegraphics[width=0.45\textwidth]{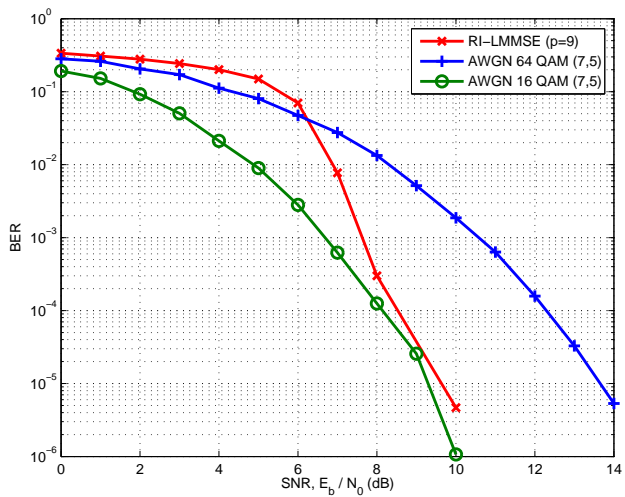}
   \caption{Performance of RI-LMMSE for FTN signaling: $16$-QAM}
   \label{fig:FTN_16qam}
\end{figure}

Fig.~\ref{fig:FTN_16qam} shows the results of the $16$-QAM signaling under $\tau=0.67$ for the same settings with BPSK signaling except that maximum $25$ of the ISI taps take part in the RI-LMMSE equalization procedure and the initial scaling factors of ($0.5,0.6$) are chosen for the LLR values of the equalizer and decoder output respectively for RI-LMMSE receiver. As seen from Fig.~\ref{fig:FTN_16qam}, the BER performance of RI-LMMSE is getting closer to the no ISI case particularly below BER of $10^{-3}$. It is also interesting to observe that the $16$-QAM FTN signaling case with these settings has the same PSD shape and transmission rate as coded ordinary $64$-QAM modulation with the same pulse shape filter. However, the performance of the FTN signaling with the proposed RI-LMMSE method has a $3-4$ dB SNR advantage with respect to the coded $64$-QAM modulation with no FTN below BER of $10^{-3}$.

 It should be noted that there is no performance characterization of the proposed methods in the literature for FTN Signaling under high order constellations within the knowlegde of the authors. One reason may be the exponentially increasing computational complexity of those methods with the constellation size. 
\section{CONCLUSION}
\label{sec:conclsn}
In this paper, we propose a graph-based low complexity LMMSE receiver structure for FTN signaling in which the statistics of colored noise is used without the necessity of a whitening filter. Our method comes to the forefront owing to its linearly increasing computational complexity with block length. Moreover, its complexity is not affected by the constellation size utilized; hence, it is very practical even when the size of the symbol alphabet is large. We also provide a new technique on extrinsic information exchange mechanism for turbo operation under the Gaussian input signal approximation. When a lower symbol time than Mazo limit is used, FTN signaling may still provide more effective transmission although this requires increased transmission power. Application of our reduced iterative LMMSE equalization method to this scenario seems to be an interesting area of future work.

\bibliography{FTN}

\end{document}